\documentclass[12pt]{iopart}

\usepackage{epsfig}
\usepackage{graphicx}
\usepackage{dcolumn}
\usepackage{bm}

\begin{document}

\title{Importance of two current model in understanding the electronic transport behavior of inverse Heusler alloy: Fe$_{2}$CoSi}
\author{Sonu Sharma and Sudhir K. Pandey}
\address{School of Engineering, Indian Institute of Technology Mandi, Kamand - 175005, India}
\address{Electronic mail: sonusharma@iitmandi.ac.in}

\date{\today}

\begin{abstract}
Here we explore the applicability of the two current model in understanding the transport behavior of Fe$_{2}$CoSi compound by using the first principles calculations in combination with the Boltzmann transport theory. The spin-unpolarized calculation shows large density of states (DOS) at Fermi level (E$_F$) and is unable to provide the correct temperature dependence of transport coefficients. The spin-polarised calculation shows reduced DOS at the E$_F$ in the spin-up channel, whereas spin-dn channel have almost zero DOS at the E$_F$. The absolute value of Seebeck coefficient in the spin-up channel shows linear increment with the temperature and in the spin-dn channel it varies non-linearly. The electrical conductivity also shows non-linear temperature dependence in both the spin channels whereas, the electronic thermal conductivity shows linear temperature dependence. The values of transport coefficients and their temperature dependence obtained by using the two current model are found to  be in fairly good agreement with the experimental data. Present work clearly suggests the importance of two current model in understanding the transport properties of the compound.
\end{abstract}


\maketitle

\section{Introduction}
In the ferromagnetic metals one can distinguish the electrons, according to the parallel or antiparallel direction of their magnetic moment to the total magnetization. The charge carriers parallel to the total magnetization are referred as spin-up electrons and antiparallel to the total magnetization are referred as spin-dn electrons. In these materials the two bands are not equally filled and hence lead to a non-zero magnetization. Mott described that the spin splitting of the energy bands in the ferromagnetic spin states induces the specific transport behavior\cite{mott}. He gave a model for the electrical conduction in ferromagnetic transition metals by considering the two independent parallel currents, carried by spin-up and spin-dn electrons. In the two current model the scattering events with the conservation of spin direction become more probable at low temperature i.e. at temperature lower than the Curie temperature ($T_{c}$). Above $T_{c}$, the application of the two-current model is limited by spin-mixing effects. 

According to this model, in the absence of any spin-flip processes the resistivities from two spin channels are added as that of two parallel resistors. The resulting average isotropic resistivity is given by

\begin{equation}
\label{eq.1}
1/\rho = 1/\rho^{\uparrow} + 1/\rho^{\downarrow}
\end{equation}
and hence the total conductivity can be expressed as\cite{dorleijn}
\begin{equation}
\label{eq.2}
\sigma = \sigma^{\uparrow}+\sigma^{\downarrow}
\end{equation}
where $\rho^{\uparrow}$ and $\rho^{\downarrow}$ are the resistivities and $\sigma^{\uparrow}$ and $\sigma^{\downarrow}$ are the conductivities of the spin-up and spin-dn channels, respectively. 

In the two current model the total Seebeck coefficient ($S$) can be expressed as the average of Seebeck coefficients from both the channels weighted by the corresponding conductivities\cite{xiang,botana}      
 
\begin{equation}
\label{eq.3}
S = [\sigma^{\uparrow}S^{\uparrow}+\sigma^{\downarrow}S^{\downarrow}]/[\sigma^{\uparrow}+\sigma^{\downarrow}]
\end{equation}

where, $S^{\uparrow}$ and $S^{\downarrow}$ are the Seebeck coefficients of the spin-up and spin-dn channels, respectively. The effective electronic thermal conductivity ($\kappa_{e}$) can be expressed as\cite{bauer}

\begin{equation}
\label{eq.3}
\kappa_{e} = \kappa_{e}^{\uparrow}+\kappa_{e}^{\downarrow}
\end{equation}

The two current model has successfully been applied to understand the transport properties of ferromagnetic transition metals\cite{dorleijn,fert,campbell} and even for antiferromagnetic transition-metal oxides\cite{botana}. As stated earlier this model is applicable well below the $T_{c}$ and compounds with higher $T_{c}$ are important for technological applications. Therefore, in the previous calculations we have applied this model to study Co$_{2}$MnGe Heusler alloy having $T_{c}$ equals to 905 K\cite{sonu}. But we did not find much importance of the model in the compounds because of the presence of the large band gap ($\sim$ 0.4 ev) in the spin-dn channel. This motivates us to search for the compound with very small spin-dn band gap so that this channel can also participate in the transport behavior of the compound. We came across the  Fe$_{2}$CoSi compound which has almost negligible band gap in the spin-dn channel. This compound is having a Curie temperature of 1038 K\cite{du}. Based on first principles calculations, Luo et al. have shown  that Fe$_{2}$CoSi is typical half-metallic ferromagnets\cite{hzluo}. However, Du et al. have proposed it as a gapless half-metallic ferromagnetic alloy\cite{du}. Half-metallic ferromagnets have attracted scientific and technological interest because of their potential in spintronic applications\cite{felser}. In these compounds only one spin channel contributes to the transport properties if the band gap of the another one is large. However, contributions from both the channels are expected when the band gap becomes almost zero. Thus in Fe$_{2}$CoSi compound both the spin channels are expected to play an important role in deciding the transport properties of this compound. Ishii et al have reported the temperature dependent data of Seebeck coefficient, electrical conductivity and electronic thermal conductivity of Fe$_{2}$CoSi above the room temperature\cite{ishii} and show significant temperature dependence. In the light of these data it will be interesting to see the applicability of the two current model in this compound.     

Here we report the electronic and transport properties of Fe$_{2}$CoSi alloy by using the electronic structures calculations and the Boltzmann transport theory. The spin-unpolarized calculation shows $\sim$ 25 states/eV/f.u. (f.u. $\equiv$ formula unit) at the \textit{E$_F$} and do not provide the correct temperature dependence of the transport coefficients. The spin-polarized calculation provides $\sim$ 1 states/eV/f.u. DOS at the \textit{E$_F$} for the spin-up channel while the spin-dn channel has almost zero DOS. In the spin-up channel the absolute value of S increases linearly with the temperature and in the spin-dn channel it varies non-linearly. In both the channels the electrical conductivity is found to vary non-linearly, whereas the electronic thermal conductivity varies linearly with the temperature. The overall temperature dependence of the transport coefficients is computed by combining the contributions from both the channels under two current model. The transport coefficients data thus obtained agree fairly well with the experimental data.

\section{Computational details and Crystal structure}
The electronic, magnetic and thermoelectric properties of the compound are studied by using the full potential linearized augmented plane-wave (FP-LAPW) method implemented in WIEN2k code\cite{blaha} in combination with the BoltzTraP code\cite{singh}. The exchange-correlation functional was taken within the generalized gradient approximation (GGA) of Perdew et al.\cite{perdew}. The muffin-tin radii were fixed to 2.25 Bohr (a$_{0}$) for Fe atom, 2.30 a$_{0}$ for Co atom and 1.90 a$_{0}$ for Si atom. $R_{MT}K_{MAX}$ was set equal to 8, where $R_{MT}$ is the smallest atomic sphere radii and $K_{max}$ is the plane wave cut-off. The self-consistency was achieved by demanding the convergence of the total energy/cell and charge/cell of the system to be less than 10$^{-5}$ Ry and 10$^{-3}$ electronic charge, respectively. The $k$-integration mesh was set to a size of ($50\times 50\times 50$) during the calculations of electronic and transport properties. The \textit{lpfac} parameter, which represents the number of k-points per lattice point was kept equal to 5 during the calculations of the transport coefficients. The value of relaxation time ($\tau$) was taken to be $5\times10^{-15} s $\cite{ashcroft}, whenever not mentioned.   

Generally, Heusler alloys prefer two kinds of structure. One is Cu$_{2}$MnAl-type structure having space group $Fm-3m$ and the other is Hg$_{2}$CuTi-type structure having space group $F-43m$. The Heusler alloys which crystallize in Hg$_{2}$CuTi-type structure are known as inverse Heusler alloys and Fe$_{2}$CoSi belongs to this. These are named so because in these alloys the atomic number of Y atom is more than the X atom of the same period i.e. if $Z(Y) > Z(X)$, opposite to that of traditional Heusler alloys. In this structure the two $Fe$ atoms occupy the $4a (0, 0, 0)$ and $4c (1/4, 1/4, 1/4)$ Wyckoff positions, while $Co$ and $Si$ atoms occupy $4b (1/2, 1/2, 1/2)$ and $4d (3/4,3/4, 3/4)$ positions, respectively\cite{graf,hluo}. In this paper Fe$_{1}$ and Fe$_{2}$ represent the Fe atoms occupying the $4a$ and $4c$ Wyckoff positions, respectively.

\section{Results and discussions}
Firstly, we will proceed with the discussion of spin-unpolarized calculation to see how far this calculation is going to help in understanding the transport behavior of the compound. The electronic band structure along the high symmetry direction in the first Brillouin zone (\textit{BZ}) is shown in the Fig. 1(a). This figure shows that Fe$_{2}$CoSi is metallic as two bands are crossing the \textit{E$_{F}$} at 10 different k-points along the W to K direction. Many bands are concentrated around the W point along the X to K direction and the system seems to be unstable. Such systems can minimize their energy by shifting these bands and this can lead to the ferromagnetic (FM) ground state as per Stoner theory\cite{skpandey,sonu}.

The total density of states (TDOS) and partial density of states (PDOS) plots are presented in Fig. 1. It is clear from Fig. 1(b) that there is very large density of states $\sim$ 25 states/eV/f.u. at the \textit{E$_{F}$}. Since in the present system the \textit{E$_{F}$} is mainly contributed by the $d$ orbitals with negligibly small contribution from the other orbitals, so we have shown the partial density of states (PDOS) of Fe and Co atoms in Fig. 1. The PDOS plot of Fe$_{1}$ atom is presented in Fig. 1(c). This plot shows that around \textit{E$_{F}$} the main contribution ($\sim$ 5 states/eV/atom) is from the $e_{g}$ states with small contribution ($\sim$ 1 states/eV/atom) from the $t_{2g}$ states. The Fe$_{2}$ atom shown in Fig. 1(d), provides more contribution than the Fe$_{1}$ atom at the \textit{E$_{F}$}. The contributions from the $e_{g}$ and $t_{2g}$ states are $\sim$ 7 and 4 states/eV/atom, respectively at the \textit{E$_{F}$}. From the PDOS of Co atom (Fig. 1(e)) it is clear that at \textit{E$_{F}$} the large contribution is from $e_{g}$ states ($\sim$ 6 states/eV/atom) and from the $t_{2g}$ states it is small. The DOS plots of Fe$_{1}$ and Co atom show three peak structures because of the presence of four fold symmetry in these atoms, whereas Fe$_{2}$ atom show mainly two peak structures as it is eight fold symmetric. From the study of PDOS plots of these atoms it is clear that the transport behavior of the compound is mainly governed by the $e_{g}$ states and the contribution from the Fe$_{2}$ atom is more.  

Now we study the temperature dependent transport behavior of the compound obtained from spin-unpolarized calculations. The variations of $S$, electrical conductivity ($\sigma/\tau$) and electronic thermal conductivity ($\kappa_{e}/\tau$) (where $\tau$ is the relaxation time) with temperature are shown in Fig. 2. The value of $S$ remains negative in the entire temperature range which implies that this compound is $n$-type. The absolute value of $S$ is found to increase with temperature upto 350 K and after that it decreases with increasing temperature as is clear from Fig. 2(a). The absolute value of $S$ at the room temperature comes out to be $\sim$ 30 $\mu$VK$^{-1}$. The value of $\sigma/\tau$ shown in Fig. 2(b), decreases with temperature and become almost constant at the higher temperature. The room temperature value of $\sigma$ is found to be $\sim$ 13,000 $\Omega^{-1}cm^{-1}$. Fig. 2(c) shows the almost linear increment in $\kappa_{e}/\tau$ with increasing temperature. The value of $\kappa_{e}$ is $\sim$ 9 $Wm^{-1}K^{-1}$. On comparing the temperature dependent behavior of the transport coefficients with the experimental results\cite{ishii} one can easily concluded that the spin-unpolarized calculation is unable to explain the temperature dependent behavior of the experimental data. 

As the spin-unpolarized transport coefficients do not show agreement with the experimental results, so we have studied the compound using the spin-polarized calculation. The spin-polarized calculation is expected to change the DOS around the \textit{E$_{F}$} by shifting the bands and consequently the transport behavior. The spin-polarized dispersion curves along the high symmetry direction of 1st \textit{BZ} are presented in the Fig. 3(a and b). The spin-up channel is found to be metallic as 3 highly dispersing bands are crossing the \textit{E$_{F}$} at 6 different k-points along the W to K direction. In the spin-dn channel one can find that the bands are not much dispersing and they slightly cross the \textit{E$_{F}$} at $\Gamma$ and $X$-points. Thus this compound show almost half-metallic ferromagnetic behavior. Also the band structure of the spin-dn channel of this inverse Heusler alloy differes from the normal Heusler alloys, where one can see the almost flat conduction band along $\Gamma$ to $X$-direction\cite{sonu,ssonu,yabuuchi,barth,tgraf,jbarth,picozzi}. The density of states plots of Fe$_{2}$CoSi are also presented in the Fig. 3. In the TDOS plot shown in Fig. 3(c) there are three peaks in the valence band of the minority and majority spin channels as is observed in case of spin-unpolarized calculation. The TDOS at the \textit{E$_{F}$} of majority spin channel is $\sim$ 1 states/eV/f.u and is nearly 25 times less than the spin-unpolarized calculation. In the spin-dn channel there is negligibly small DOS at the \textit{E$_{F}$} so this channel show almost semiconducting behavior. Similar TDOS plot for this compound is also observed in the earlier theoretical work\cite{du,hzluo,hluo}. 

To see the contributions from various atomic states around the \textit{E$_{F}$}, we have presented PDOS plots in Fig. 3(d-f). In these plots only $d$-orbitals are shown for the Fe$_{1}$, Fe$_{2}$ and Co atoms. The spin-up channel of Fe$_{1}$ and Co atoms is contributed by both $t_{2g}$ and $e_{g}$ states, with the value less than 0.5 states/eV/atom at the \textit{E$_{F}$}. In the spin-dn channel, the \textit{E$_{F}$} is at the edge of the $t_{2g}$ states of Fe$_{1}$ and Co atoms and shows the almost semiconducting behavior. The PDOS plot of Fe$_{2}$ atom shows only two peak structure. In the majority spin channel there are two peaks below the \textit{E$_{F}$} and in the minority spin channel one peak is below and other is above the \textit{E$_{F}$} due to the exchange splitting. It is clear from these plots that, mainly the $t_{2g}$ states of Fe$_{1}$ and Co atoms are contributing to the DOS around the \textit{E$_{F}$} in the minority spin channel. Thus in Fe$_{2}$CoSi, the contribution in the transport properties is expected from both the spin channels. Also around the \textit{E$_{F}$} different states are contributing to the TDOS of both the channels so the relaxation time ($\tau$) is also expected to be different for the both channels.  

The value of magnetic moment of the compound comes out to be $\sim$ 5 $\mu_{B}$ which is as per the Slater-Pauling rule\cite{galanakis,graf} and is closer to the experimental values \cite{hzluo,du}. The contribution from the Fe$_{1}$, Fe$_{2}$ and Co atoms is $\sim$ 1.41, 2.69 and 1.05 $\mu_{B}$, respectively. The maximum contribution to the total magnetic moment is from the Fe$_{2}$ atom because of large exchange splitting in the $3d$ orbitals of Fe$_{2}$ atom. 

The DOS plots of the compound show the contribution from different states around the \textit{E$_{F}$}, so the transport behavior is also expected to be different for both the channels. Therefore we have studied the transport coefficients of the compound for both the spin channels and transport coefficients versus temperature plots are presented in Fig. 4. The Seebeck coefficient for spin-up and spin-dn channels are shown in Fig. 4(a) and 4(b), respectively. The $S$ for spin-up channel is negative and its absolute value increases linearly with temperature. The room temperature value of $S_{\uparrow}$ is $\sim$ -11 $\mu$VK$^{-1}$. The negative value of $S$ indicates that this is $n$-type thermoelectric material. For the spin-dn channel the variation of $S$ with temperature is not linear, its absolute value firstly increases upto 300 K and then start decreasing upto 500 K. After 500 K the $S$ becomes positive and again starts increasing with the temperature. The value of $S_{\downarrow}$ at room temperature is around -13 $\mu$VK$^{-1}$. Experimentally, the absolute value of $S$ shows non linear increment with the temperature, which can only be understood if one combines the contributions from both the channels.  

Also we have plotted the electrical conductivity with temperature for both the spin channels in Fig. 4(c and d). For the spin-up channel the $\sigma/\tau$ decreases with increasing temperature which corresponds to the metallic behavior. The value of $\sigma$ at 300 K is about 38,000 $\Omega^{-1}cm^{-1}$. In the spin-dn channel it increases with the temperature like semiconductors. The spin-dn channel gives very less conductivity, clearly because of low value of DOS at the \textit{E$_{F}$}. Both these channels show non linear variation with the temperature and have curvatures in opposite directions. However, in experimental plots the $\sigma$ decreases almost linearly with the temperature, thus one can expect the overall combination from both the channels can give such temperature dependent behavior of $\sigma$.

Fig. 4(e and f) show the plots of electronic thermal conductivity with the temperature for both the spin channels. The $\kappa_{e}/\tau$ increases linearly with the temperature in the spin-up channel and at 300 K it is $\sim$ 28 $Wm^{-1}K^{-1}$. The spin-dn channel shows that $\kappa_{e}/\tau$ increases non-linearly with the temperature as there is a curvature. In the spin-dn channel $\kappa_{e}$ is almost 10 times smaller and its value at the room temperature is $\sim$ 1 $Wm^{-1}K^{-1}$. 

From the above discussions it is clear that the overall temperature dependent behavior of transport coefficients can be understood by combining the contributions from both the channels. Therefore we have applied the two current model to find the total Seebeck coefficient, electrical conductivity and electronic thermal conductivity of the system. The total $S$, $\sigma$ and $\kappa_{e}$ are presented in the Fig. 5(a-c). The value of total $S$ is negative and decreases with the temperature. The variation of S with temperature using the two current model shows non linear temperature dependence, as is observed experimentally. The absolute value of S at room temperature is $\sim$ 11 $\mu$VK$^{-1}$ and is closer to the experimental value of $\sim$ 13 $\mu$VK$^{-1}$. The $\sigma$ shows non linear increment with the temperature while $\kappa_{e}$ increases linearly. The temperature dependent behavior of the $\sigma$ and $\kappa_{e}$ does not show agreement with the experimental data. This may be due to the reason that the value of the relaxation time has been taken same for the both spins. As mentioned earlier, the different states are contributing around the \textit{E$_{F}$} therefore one expects different values of $\tau$ for both the spin channels. 

If we take the values of $\tau$ equal to $5\times 10^{-15}$ and $2\times 10^{-15}$ s for spin-up and spin-dn channels, respectively the calculated values of $\sigma$ and $\kappa_{e}$ are found to be in fairly good agreement with the experimental data as evident from Fig. 6. The room temperature value of $\sigma$ agrees with the experimental value of 19,000 $\Omega^{-1}cm^{-1}$. The plot of the $\kappa_{e}$ versus temperature is found to increase almost linearly with the temperature. The value of $\kappa_{e}$ at the room temperature is $\sim$14 $Wm^{-1}K^{-1}$ which is comparable with the experimental value. At this stage it is important to note that the experimental $\kappa_{e}$ has been calculated by using the Wiedemann-Franz law and it shows a curvature at the higher temperature. However, calculated $\kappa_{e}$ shows almost linear dependence. This deviation suggests the inadequacy of the Wiedemann-Franz law in understanding the temperature dependent behavior of the $\kappa_{e}$. Above results clearly show the importance of two current model in understanding the temperature dependent behavior of the transport coefficients. It is also important to note that we have not considered the temperature dependent $\tau$, thus the values of $\sigma$ and $\kappa_{e}$ are expected to be different at the higher temperature.

\section{Conclusions}
The electronic and transport properties of the Fe$_{2}$CoSi inverse Heusler alloy are studied by combining the electronic band structure calculations with the Boltzmann transport theory. The spin-unpolarized calculation shows a very large DOS at the \textit{E$_F$} and is unable to predict the correct temperature dependent behavior of the transport coefficients. The spin-polarized calculation provides $\sim$ 25 times less DOS at the \textit{E$_F$} for the spin-up channel whereas there is almost zero DOS in the spin-dn channel. Fe atom at 4c Wyckoff position shows large exchange splitting and contribute largely to the total magnetic moment. The S$^{\uparrow}$ varies linearly while S$^{\downarrow}$ shows non linear relation with the temperature. The $(\sigma/\tau)^{\uparrow}$ decreases while $(\sigma/\tau)^{\downarrow}$ increases non linearly with the temperature. The electronic thermal conductivity also found to increase with the increasing temperature for both the channels. The experimentally observed temperature dependent behavior of the compound is well explained by combining the transport coefficients of both the channels and  considering different values of relaxation time. The room temperature values of S, $\sigma$ and $\kappa_{e}$ are 11 $\mu$VK$^{-1}$, 19,000 $\Omega^{-1}cm^{-1}$ and 14 $Wm^{-1}K^{-1}$, respectively and show good agreement with the experimental values.

\section{Figures}

\begin{figure}
  \includegraphics[width=14cm]{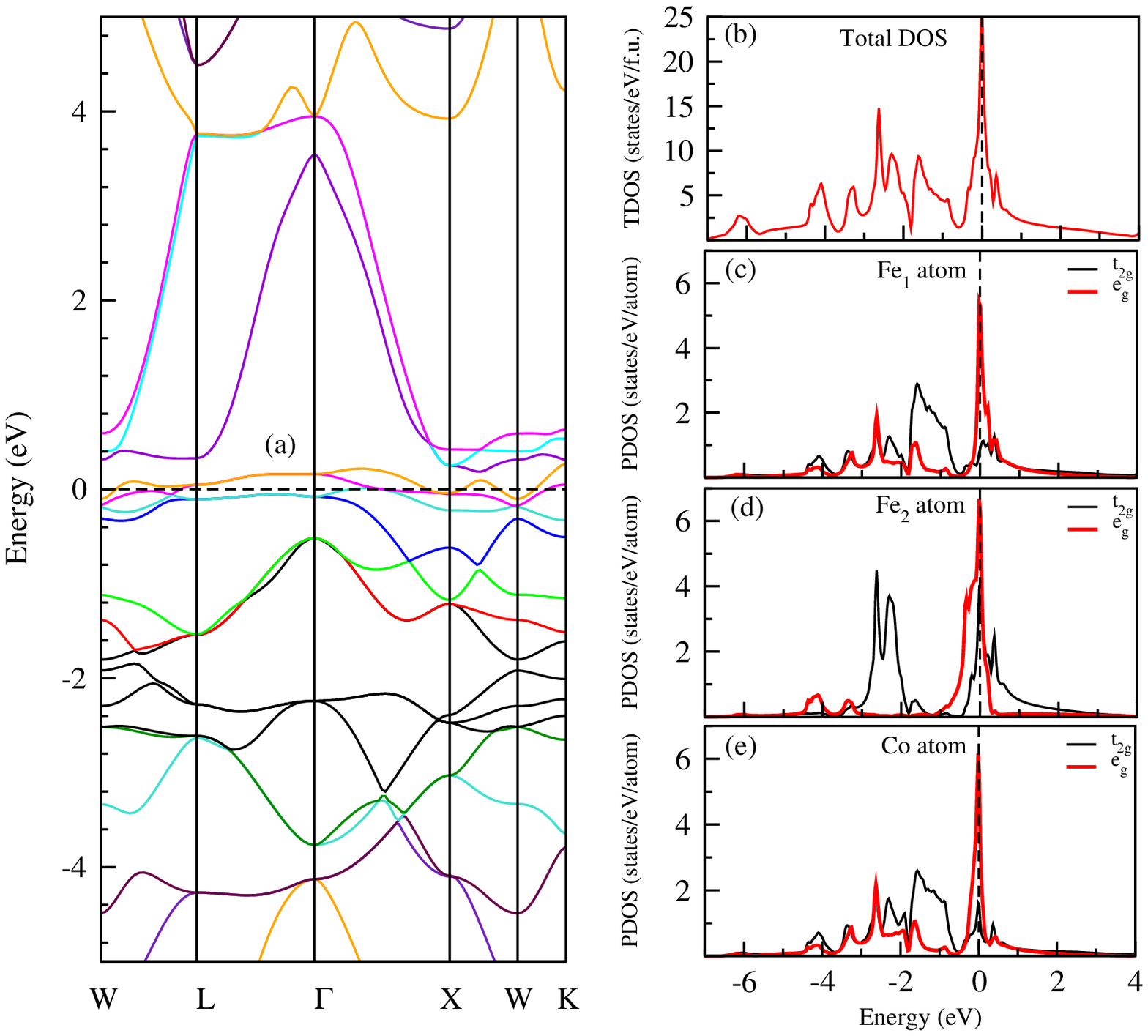}\\
  \caption{(Color online) (a) Electronic band structure, (b) TDOS, (c) PDOS of Fe$_{1}$ atom (Fe at 4a wyckoff position), (d) PDOS of Fe$_{2}$ atom (Fe at 4c wyckoff position) and (e) PDOS of Co atom for Fe$_{2}$CoSi compound in spin-unpolarized calculation.}\label{Fig1}
\end{figure}

\begin{figure}
  \includegraphics[width=14cm]{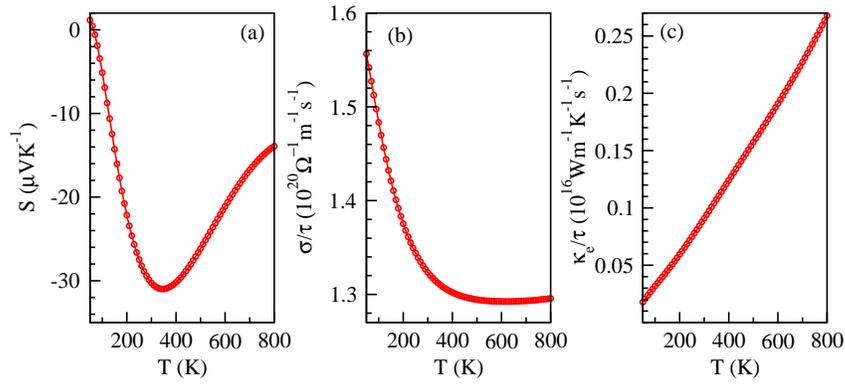}\\
  \caption{(Color online) Temperature variation of (a) Seebeck coefficient, (b) electrical conductivity and (c) electronic thermal conductivity with temperature for Fe$_{2}$CoSi compound in spin-unpolarized calculation.}\label{Fig2}
\end{figure}

\begin{figure}
  \includegraphics[width=14cm]{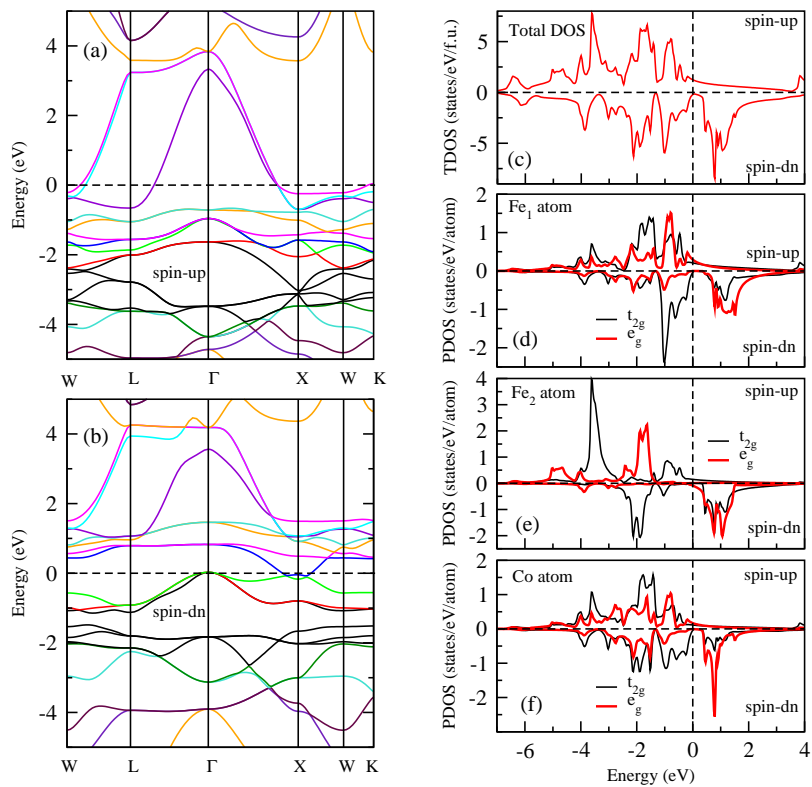}\\
  \caption{(Color online) Spin-polarized electronic band structures and Total and partial density of states plots of Fe$_{2}$CoSi. Shown are (a) band structure of spin-up channel, (b) band structure of spin-down channel, (c) TDOS, (d) PDOS of Fe$_{1}$ atom (Fe at $4a$ wyckoff position), (e) PDOS of Fe$_{2}$ atom (Fe at $4c$ wyckoff position) and (f) PDOS of Co atom.}\label{Fig3}
\end{figure}

\begin{figure}
  \includegraphics[width=14cm]{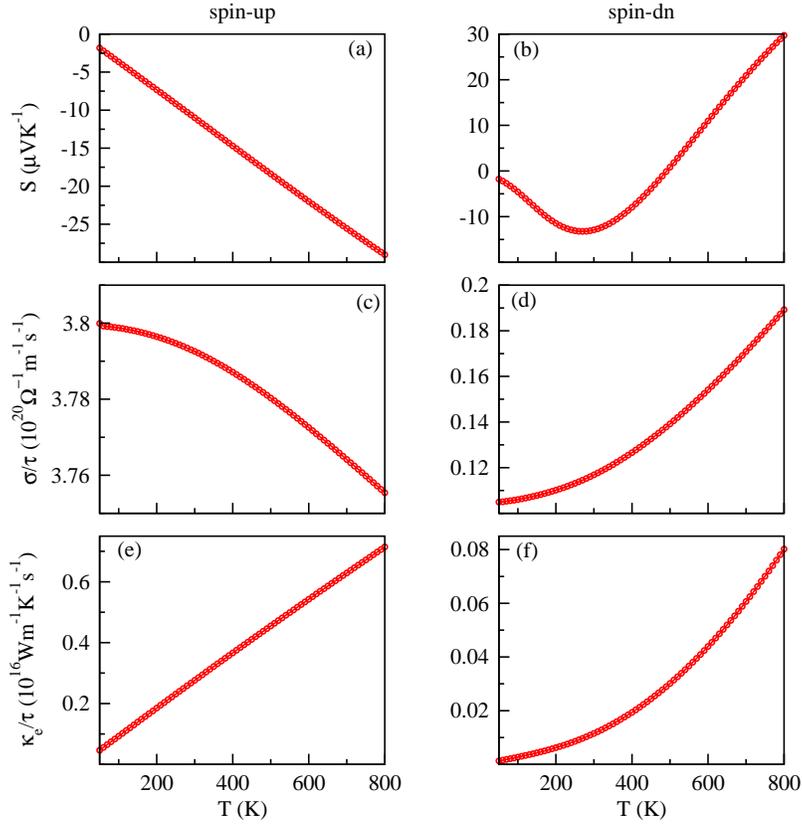}\\
  \caption{(Color online) Variation of transport coefficients with temperature. (a and b) Seebeck coefficient with temperature, (c and d) electrical conductivity with temperature and (e and f) electronic thermal conductivity with temperature.}\label{Fig4}
\end{figure}

\begin{figure}
  \includegraphics[width=14cm]{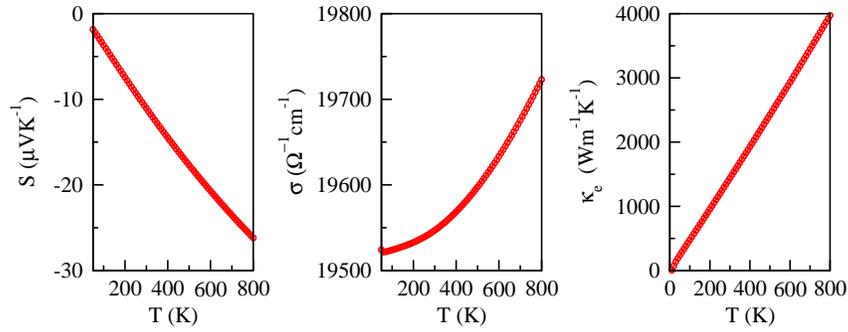}\\
  \caption{(Color online) Variation of (a) Seeebeck coefficient, (b) electrical conductivity and (c) electronic thermal conductivity with temperature within the two current model (relaxation time ($\tau$) is taken to be 5$\times10^{-15}$ s for both the spin channels).}\label{Fig5}
\end{figure}

\begin{figure}
  \includegraphics[width=14cm]{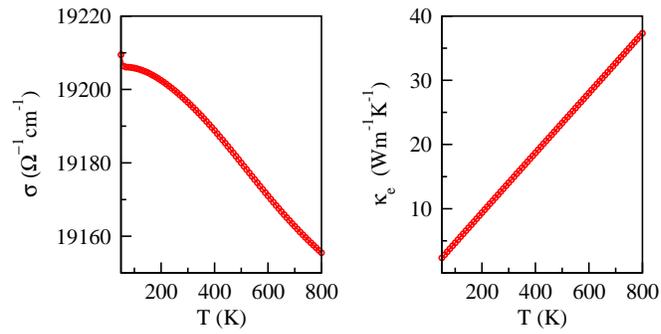}\\
  \caption{(Color online) Variation of (a) electrical conductivity and (b) electronic thermal conductivity with temperature within the two current model ($\tau$ is taken to be 2 $\times10^{-15}$ and 5 $\times10^{-15}$ s for spin-up and spin-dn channels, respectively).}\label{Fig6}
\end{figure}

\end{document}